\documentclass[%
 reprint,
 amsmath,amssymb,
 aps,
]{revtex4-2}
\usepackage{cprotect}
\usepackage{xcolor} 
\usepackage{graphicx}
\usepackage{dcolumn}
\usepackage{bm}
\usepackage{float}
\usepackage{setspace}
\usepackage{subcaption}
\usepackage{comment}
\usepackage{adjustbox}
\usepackage{multirow}
\newcommand{\boroneight}{$^{8}$B}

\begin{document}

\preprint{APS/123-QED}

\title{Slow-Fluor Scintillator for Low Energy Solar Neutrinos \\ and Neutrinoless Double Beta Decay}

\author{Jack Dunger}
\author{Edward J. Leming}
\author{Steven D. Biller}
\affiliation{%
 University of Oxford \\
 Denys Wilkinson Building, Keble Road, Oxford, OX1 3RH, UK \\ \\
 ({\it Physical Review D, Volume 105, No. 9, 1 May 2022})
}%


\begin{abstract}
The potential for using slow-fluor liquid scintillators to study low energy solar neutrinos and neutrinoless double beta decay ($0\nu\beta\beta$) is explored through a series of simulations. The fluorescence model assumed for the primary fluor has characteristics similar to acenaphthene, recently used to demonstrate Cherenkov separation at energies around 1~MeV \cite{SlowFluors}. Results here indicate notably better directional reconstruction in large-scale detectors than has previously been suggested by other approaches, allowing better identification of low energy solar neutrinos. These studies indicate that a detector with as little as $\sim$30\% coverage using currently available photomultiplier tubes could be able to make a measurement of the CNO solar neutrino flux to a precision of better than 10\% (enough to distinguish metallicity models) with a few kiloton-years of exposure. In terms of $0\nu\beta\beta$, studies here suggest that the ability to separate mechanisms based on angular distributions is weak, but that the rejection of solar neutrino backgrounds with such a technique might potentially approach a factor of 10 for endpoint energies near 2.5 MeV in the angular hemisphere defined by the solar direction.


\end{abstract}

\maketitle

\section{\label{sec:introduction} Introduction}
A practical approach to the efficient separation of Cherenkov and scintillation components using slow fluors in liquid scintillation detectors has recently been demonstrated on a bench-top scale \cite{SlowFluors}. This approach achieves an effective time-separation of the prompt Cherenkov signal by using scintillator formulations with high light yield but slow fluorescence, permitting topological and directional information to be obtained while still maintaining good energy resolution. In this paper, the potential applications of this technique to low energy solar neutrinos and neutrinoless double beta decay ($0\nu\beta\beta$) in large scale liquid scintillation detectors are examined through a series of simulation studies. This work differs from some previous studies (such as \cite{Land}) in several important aspects, including a more realistic treatment of scintillator optical and timing characteristics, using measured properties of a known slow-fluor scintillator; a consideration of the impact of secondary fluor concentrations; a more practical detector scale and configuration with levels of photocathode coverage that have been achieved in large scale detectors; and photodetectors with timing characteristics comparable to what is currently available. The results indicate that notably better directional information and greater sensitivity for low energy solar neutrinos might be achieved using current technology in a modest scale detector than has previously been suggested.

As the purpose of this current work to explore what might be achieved subject to different practical detector requirements, this will be done in the context of a generic detector using photomultipliers (PMTs) with both faster and slower time responses and operating at two different overall detection efficiencies (the product of PMT efficiency and coverage). The impact of secondary fluors is also considered. In this study, liquid scintillator with a light yield and timing characteristics similar to acenaphthene scintillator \cite{SlowFluors} will be assumed. Acenaphthene seems to be a particularly good primary fluor for this due to its long fluorescence decay time ($\sim$45ns) and reduced absorption at lower wavelengths compared with PPO, which is favourable for Cherenkov light. It should be noted that this may not be the optimum choice for $0\nu\beta\beta$ experiments using loaded scintillator, as the light output of this fluor is $\sim$35\% lower than PPO and is particularly sensitive to quenching effects. The best choice of fluor will therefore depend on the dominant backgrounds and the specific technique used to load the target $0\nu\beta\beta$ isotope into the scintillator mixture. Nevertheless, acenapthene is used here to explore the in-principle characteristics of such an approach.

\section{\label{sec:model} Model}
\subsection{\label{sec:configuration} Configuration}

Simulations were performed using the SNO+ version of the GEANT4-based RAT software package \cite{rat}. It contains a GLG4Sim \cite{glg4sim} simulation that handles the production of scintillation photons and a full Geant4 \cite{geant4} detector simulation that individually tracks each photon through the geometry, accounting for reflection, refraction, scattering, absorption and re-emission. 

The model assumes a spherical
detector, with the target liquid scintillator housed in a spherical acrylic vessel (AV)  8.8m in radius and 5cm in thickness. Photons produced in the scintillator propagate outwards towards 21873 inward-pointing photomultiplier tubes (PMTs). Each PMT is modelled as an 8" Hamamatsu R5912 with an HQE photocathode, equipped with a reflective conical concentrator of diameter 28cm to provide an overall effective photocathode coverage of 77\%. The PMTs and concentrators are modelled as 3D objects and the front-end and trigger system are simulated in full, including electronics noise. Water fills the space between the AV and the PMTs, as well as the space behind them. The data acquisition system is the same as the one on the SNO+ detector \cite{Boger:1999bb}, but with the trigger thresholds lowered to account for the higher light detection levels. Radioactive decays are generated using a version of Decay0 \cite{Ponkratenko:2000um}. 

In this detector configuration, event energies up to a few MeV result in low photon occupancies in the PMTs, and the observed number of PMT hits is therefore approximately proportional to the energy deposited for a given event position in the detector. For the purposes of this study, as a further simplification to geometric corrections, events are only simulated in the central 1m of the detector, although fully reconstructed vertex positions are still used for the analyses that follow.

\subsection{\label{sec:baseline_and_variant} Baseline and Variants}
Four variants on the base detector configuration with different photocathode coverage and PMT speeds are considered, summarized in table~\ref{tab:configurations}. In order to achieve the configurations with the lower effective lower coverage of 30\%, the collection efficiency of the PMT model was reduced. The resulting overall photon detection efficiency is comparable to current generation experiments.
 Furthermore, we consider these configurations both with and without the addition of the secondary shifter bis-MSB, as discussed below.

\begin{table*}[t]
    \centering
    \begin{tabular}{c c c c c}
             & \% Photocathode & PMT & bis-MSB & Resulting \\
     Acronym & coverage & TTS (ns) & (mg/L)  & pe/MeV \\
     \hline
         \verb|77_FAST_1| & 77 & 1 & 1 & 1000\\
         \verb|77_SLOW_1| & 77 & 3.7 & 1 & 1000\\
         \verb|77_SLOW_0| & 77 & 3.7 & 0 & 500\\
         \verb|30_FAST_1| & 30 & 1 & 1 & 400  \\
         \verb|30_SLOW_1| & 30 & 3.7 & 1 & 400\\
         \verb|30_FAST_0| & 30 & 1 & 0 & 200 \\
         \verb|30_SLOW_0| & 30 & 3.7 & 0 & 200\\
    \end{tabular}
    \caption{Effective photocathode coverage and FWHM Transit Time Spread (TTS) for the PMTs in the four detector configurations considered.}
    \label{tab:configurations}
\end{table*}

\subsection{Scintillator}

\begin{figure}[H]
        \begin{tabular}{c}
            \includegraphics[width=7cm]{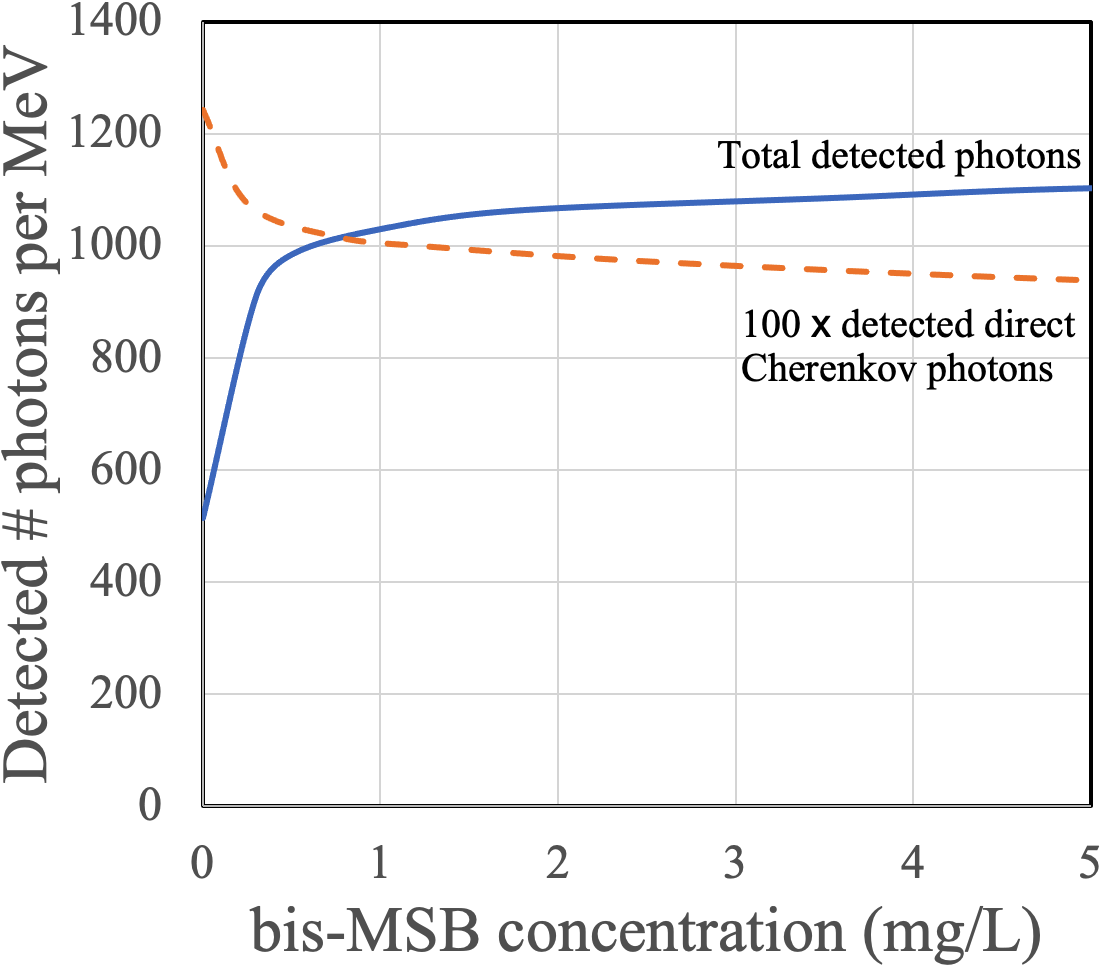} \\
        \end{tabular}
      \cprotect\caption{Average number of detected photons and 100 times the number of direct Cherenkov photons as a function of bis-MSB concentration for nominal detector configuration of this study. Curves are derived from simulations of 1 MeV electrons using fast HQE PMTs and a scintillator comprised of LAB with 4g/L of acenapthene.}
    \label{fig:bismsb_opt}
\end{figure}

The central scintillator target is modelled as Linear Alkyl Benzene (LAB) with 4g/L of acenapthene as a primary fluor \cite{SlowFluors}. In many of the configurations explored, a small amount of 1,4-Bis(2-methylstyryl)benzene (bis-MSB) is also used as a secondary fluor to increase the detected light by shifting emission to higher wavelengths to avoid LAB absorption. This is common practice in larger scale scintillation detectors (see for example \cite{JUNO}). However, the use of bis-MSB can also shift and, hence, reduce the directionality of Cherenkov light at lower wavelengths as well as impacting the balance between Cherenkov and scintillation light levels. This is indicated in Figure~\ref{fig:bismsb_opt}, which shows the relative scintillator light yield and number of `direct' ({\em i.e.} unabsorbed) Cherenkov photons per MeV of deposited energy for our nominal detector configuration. The time profile of observed light is also altered depending on the extent to which absorption and re-emission by the primary fluor is allowed by the presence of the secondary fluor. The optimal concentration therefore needs to be tailored towards particular physics goals and will ultimately depend on aspects such as light collection in the region of interest, the size of the detector, the required energy resolution and the required directionality of the detected signal. This work examines the cases of no bis-MSB and 1 mg/L bis-MSB.

\section{\label{sec:time_separation} Vertex and Direction Reconstruction}

\begin{figure}[H]
    \resizebox{!}{.35\textwidth}{
        \begin{tabular}{c}
            \includegraphics{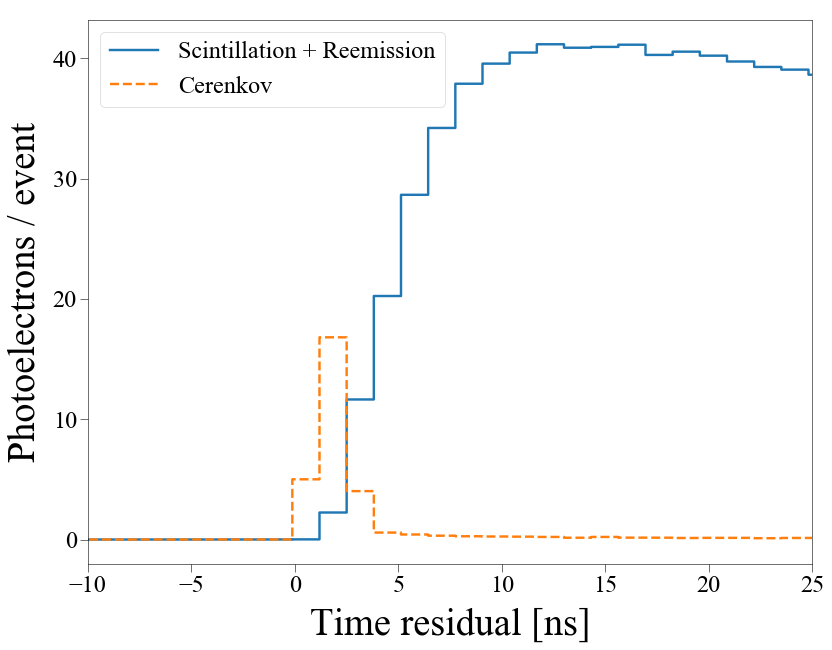} \\
            \includegraphics{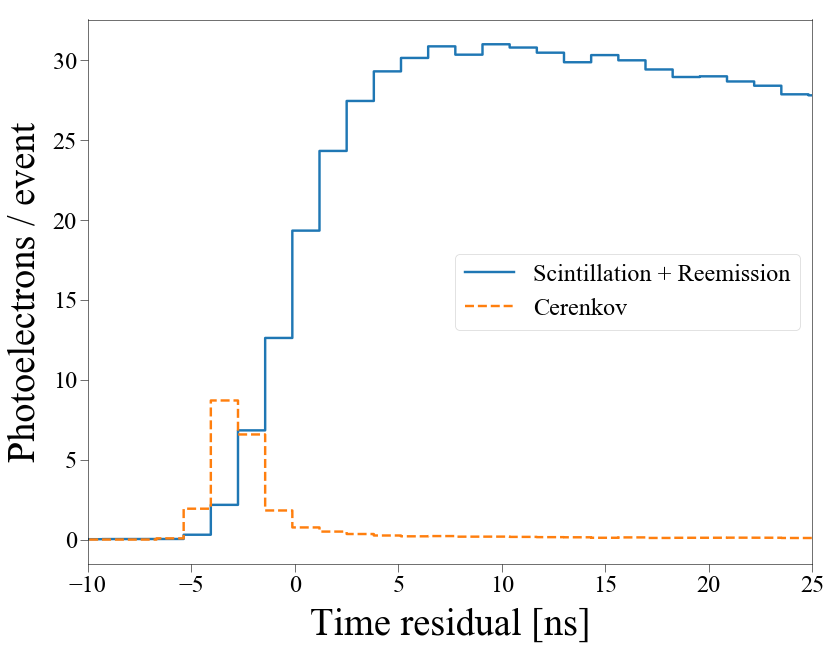}
        \end{tabular}
    }
	\cprotect\caption{Time residual spectra for 2.5 MeV electrons in the central 1~m of detector configuration  \verb|77_FAST_1|. Top: using the MC vertex positions. Bottom: using the reconstructed vertex positions.}
	\label{fig:LightSeperation}
\end{figure}

Figure~\ref{fig:LightSeperation} (top) shows the scintillation and Cherenkov hit times for 2.5 MeV electrons simulated in the central 1m of detector configuration \verb|77_FAST_1|, corrected for time of flight from the true event vertex. These `time residuals' demonstrate that the time separation demonstrated on the bench top in \cite{SlowFluors} is largely maintained through propagation in a 10m-scale scintillation detector. The early cherenkov hits are directional and can be used to reconstruct the electron direction. In practice, the vertex and direction must be estimated jointly, since identification the early hits cannot take place without knowing the vertex, and an unbiased estimate of the vertex cannot take place without knowing the direction (owing to the Cherenkov light).

The algorithm used in this study follows a maximum likelihood approach, simultaneously estimating the time, position and direction of events. Both isotropic scintillation and directional Cherenkov light are taken into account through 2D PDFs in PMT hit time and angle with respect to primary particle direction.  Owing to dispersion effects, an `effective' speed of light averaged over the detected wavelengths is chosen so as to minimise bias. The algorithm is separately optimised for simulated electrons in each of the configurations explored.

Figure~\ref{fig:LightSeperation} (bottom) shows the importance of explicitly using the fitted vertex position in evaluating the performance of time-based Cherenkov separation, as the extra degrees of freedom tend to lead to a reduction in this apparent separation.

\begin{figure}[H]
    \resizebox{!}{.27\textwidth}{
        \begin{tabular}{c}
            \includegraphics{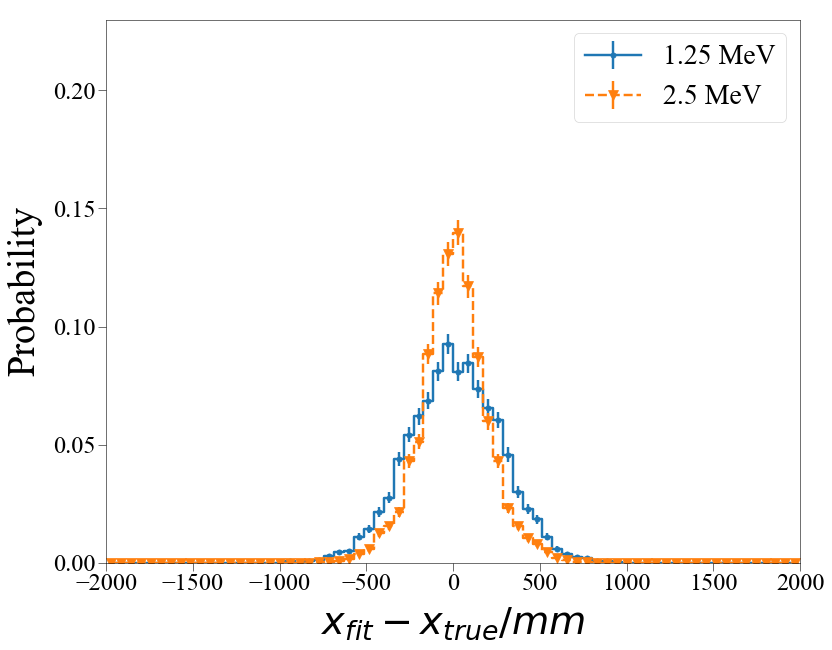} \\
            \includegraphics{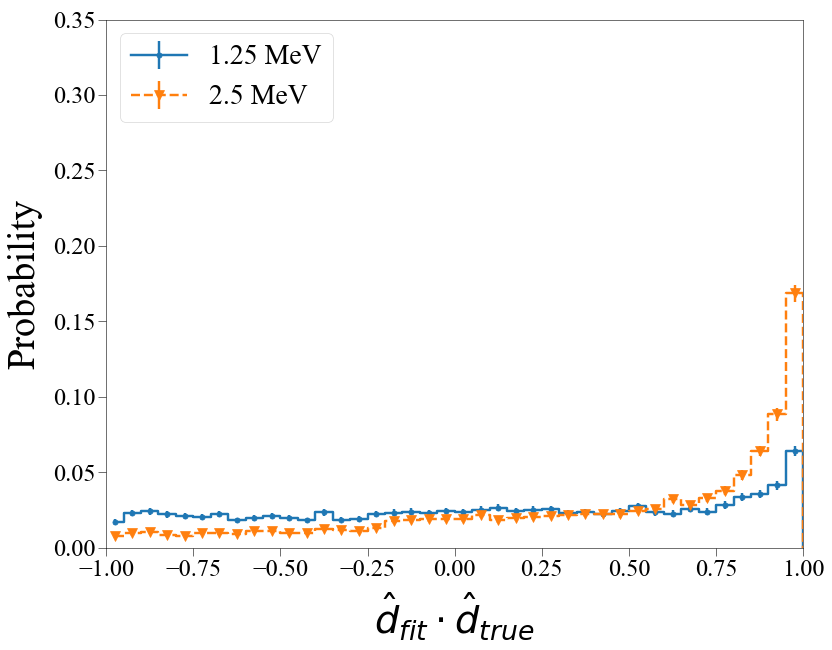} 
        \end{tabular}
    }
    \cprotect\caption{Vertex (x-projection) and direction resolution of electrons for configuration  \verb|30_FAST_1|. Blue solid: 1.25 MeV; Orange dashed: 2.5 MeV.}
    \label{fig:reconstruction_400_fast}
\end{figure}

\begin{figure}[H]
    \resizebox{!}{.27\textwidth}{
        \begin{tabular}{c}
            \includegraphics{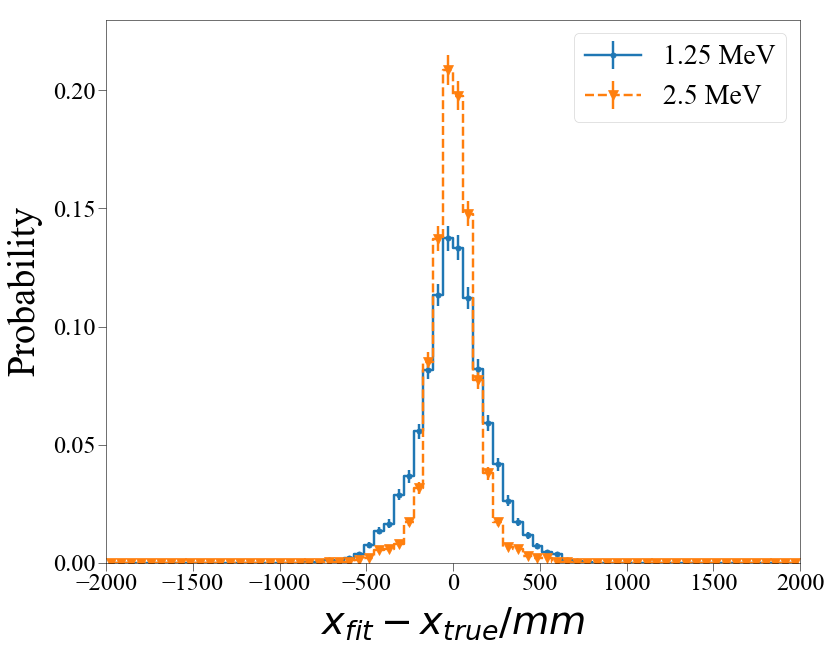} \\
            \includegraphics{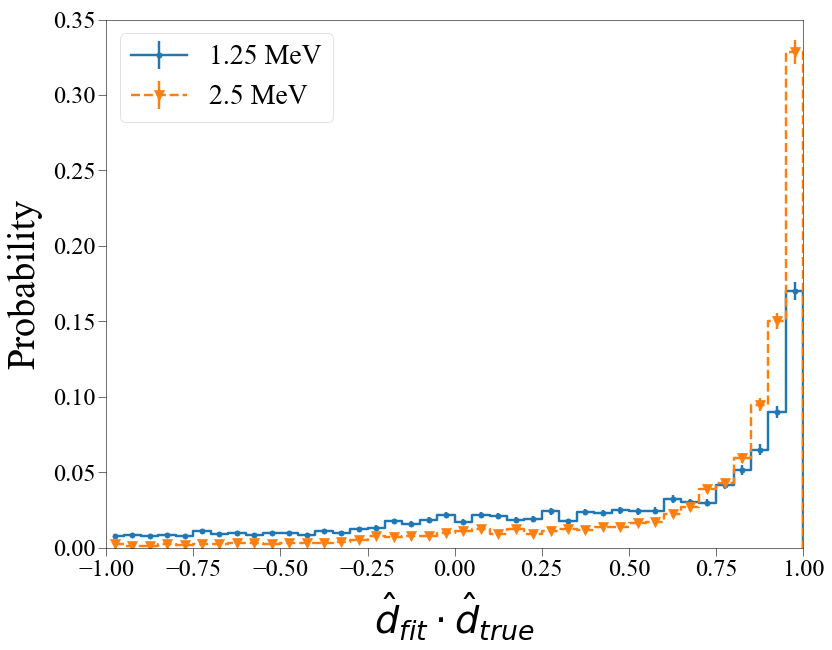} 
        \end{tabular}
    }
    \cprotect\caption{Vertex (x-projection) and direction resolution of electrons for configuration  \verb|77_FAST_1|. Blue solid: 1.25 MeV; Orange dashed: 2.5 MeV.}
    \label{fig:reconstruction_1000_slow}
\end{figure}

\begin{figure}[H]
    \resizebox{!}{.27\textwidth}{
        \begin{tabular}{c}
            \includegraphics{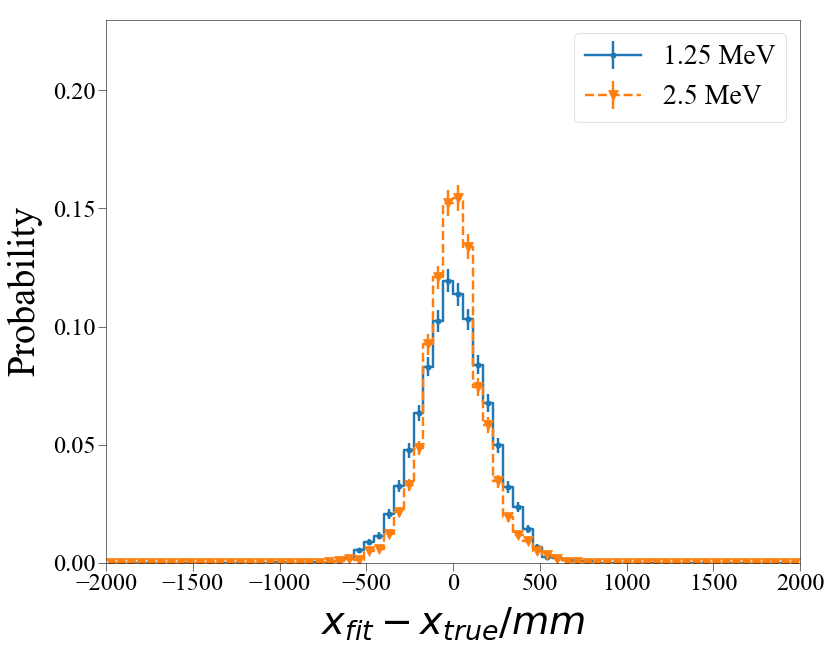} \\
            \includegraphics{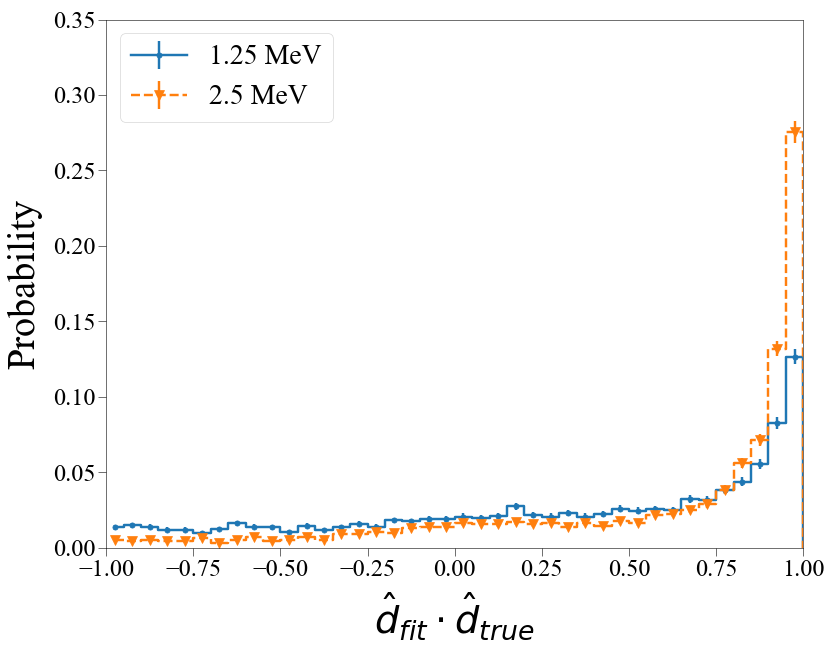} 
        \end{tabular}
    }
    \cprotect\caption{Vertex (x-projection) and direction resolution of electrons for configuration \verb|77_SLOW_1|. Blue solid: 1.25 MeV; Orange dashed: 2.5 MeV.}
    \label{fig:reconstruction_1000_fast}
\end{figure}

For the case where 1mg/L of bis-MSB is added, figure~\ref{fig:reconstruction_400_fast} indicates the performance of vertex and direction reconstruction for configuration \verb|33_FAST_1| for electrons with energies 1.25 and 2.5 MeV (of most relevance for CNO neutrinos and $0\nu \beta \beta$, respectively). Figures~\ref{fig:reconstruction_1000_slow} and \ref{fig:reconstruction_1000_fast} compare this performance for configurations \verb|77_FAST_1| and \verb|77_SLOW_1| at electron energies of 1.25 and 2.5 MeV. While the use of fast PMTs improves the vertex resolution by $\sim$30-50\%, the choice has less impact on the directional reconstruction. This is as expected since the scintillation time constant is significantly larger than the response of either type of PMT, permitting good separation of the Cherenkov component in either case.

\begin{figure}[H]
    \resizebox{!}{.27\textwidth}{
        \begin{tabular}{c}
            \includegraphics{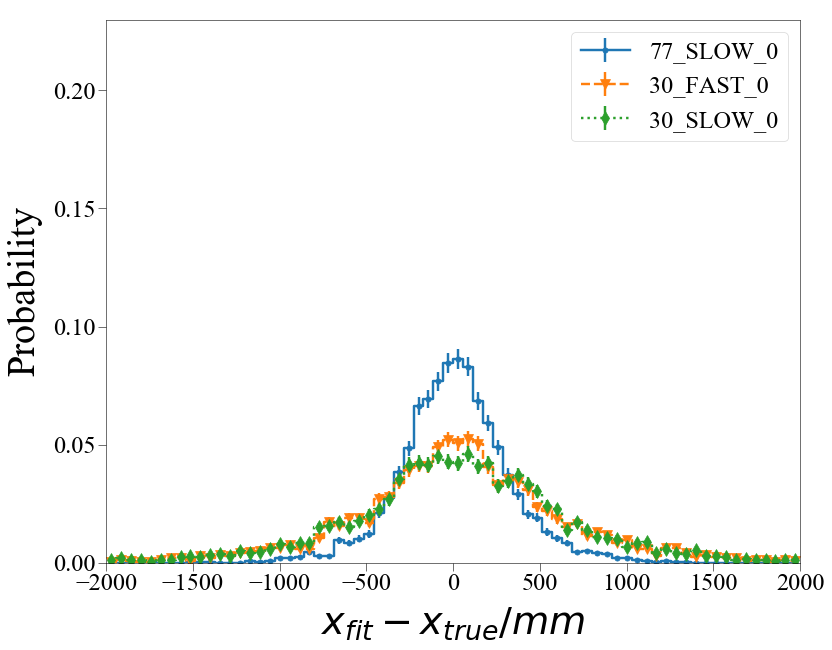} \\
            \includegraphics{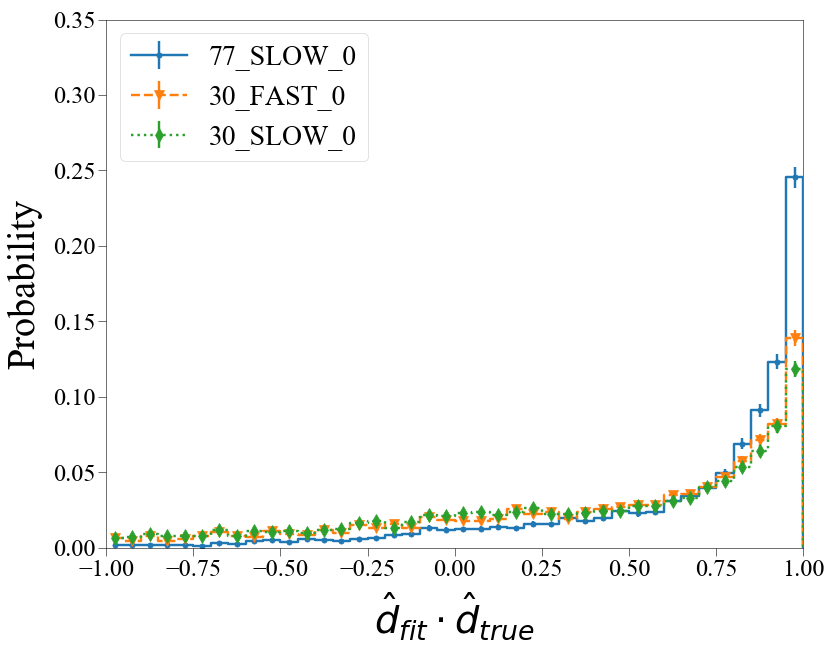} 
        \end{tabular}
    }
    \cprotect\caption{Vertex (x-projection) and direction resolution of 1.25 MeV electrons for various configurations without bis-MSB.}
    \label{fig:reconstruction_noMSB}
\end{figure}

The case of no bis-MSB is of relevance to low energy solar neutrinos, where directional separation from isotropic internal backgrounds such as $^{210}$Bi is paramount, allowing for some compromises in energy and vertex resolution. Figure~\ref{fig:reconstruction_noMSB} therefore compares the vertex and directional reconstruction for electrons with an energy of 1.25 MeV (relevant for CNO neutrinos) for detector configurations \verb|77_SLOW_0|,  \verb|30_FAST_0| and \verb|30_SLOW_0|. Results are summarized in Table \ref{tab:performance}. As can be seen, it would appear there is a significant advantage to be gained with the no bis-MSB case in terms of enhanced directional reconstruction while still preserving adequate vertex and energy resolution. Once more, the impact of PMTs with faster vs slower timing characteristics is more modest.

\begin{table*}[t]
    \centering
    \begin{tabular}{c c c c c}
     Configuration & Energy& $\sigma_x$ Position & Direction Cosine \\
     Acronym & (MeV) & Resolution (cm)& Containing 50\% \\
     \hline
         \verb|77_FAST_1| & 1.25 & 14.9 & 0.60 \\
         \verb|77_FAST_1| & 2.5 & 10.2 & 0.92 \\
         \verb|77_SLOW_1| & 1.25 & 18.7 & 0.45\\
         \verb|77_SLOW_1| & 2.5 & 12.8 & 0.84 \\
         \verb|77_SLOW_0| & 1.25 & 24.2 & 0.84 \\
         \verb|30_FAST_1| & 1.25 & 26.0 & 0.13 \\
         \verb|30_FAST_1| & 2.5 & 14.9 & 0.64 \\
         \verb|30_FAST_0| & 1.25 & 38.3 & 0.63 \\
         \verb|30_SLOW_0| & 1.25 & 42.6 & 0.55\\
    \end{tabular}
    \caption{Summary position and direction resolution performance.}
    \label{tab:performance}
\end{table*}

\section{\label{sec:solar} Solar Neutrinos}

$^{210}$Bi, supported by $^{210}$Pb, is particularly problematic for the detection of CNO neutrinos because its beta decay produces a single electron with an endpoint of 1.162 MeV. This closely mimics the CNO signal and leads to fit degeneracies. The Borexino experiment was able to measure the CNO flux to $\sim$30\% by choosing a thermodynamically stable region of their detector where mixing was sufficiently small that assumptions of equilibrium could be used to constrain this background \cite{Borexino_Nature}. In principle, slow scintillators can break the $^{210}$Bi degeneracy and reduce their background contribution much more effectively by using the fact that the reconstructed direction for neutrino events will tend to point towards the sun, whereas $^{210}$Bi events (as well of those from other radioactive backgrounds) will not. Borexino were able to demonstrate this Cherenkov directionality in a statistical sense using the earliest hit times of $\sim$10000 $^7Be$ events \cite{Borexino_direction}. The approach here would allow for a much more powerful discriminant at the event-by-event level.

\begin{figure}[H]
    \centering
    \begin{tabular}{c}
    \includegraphics[width=7cm]{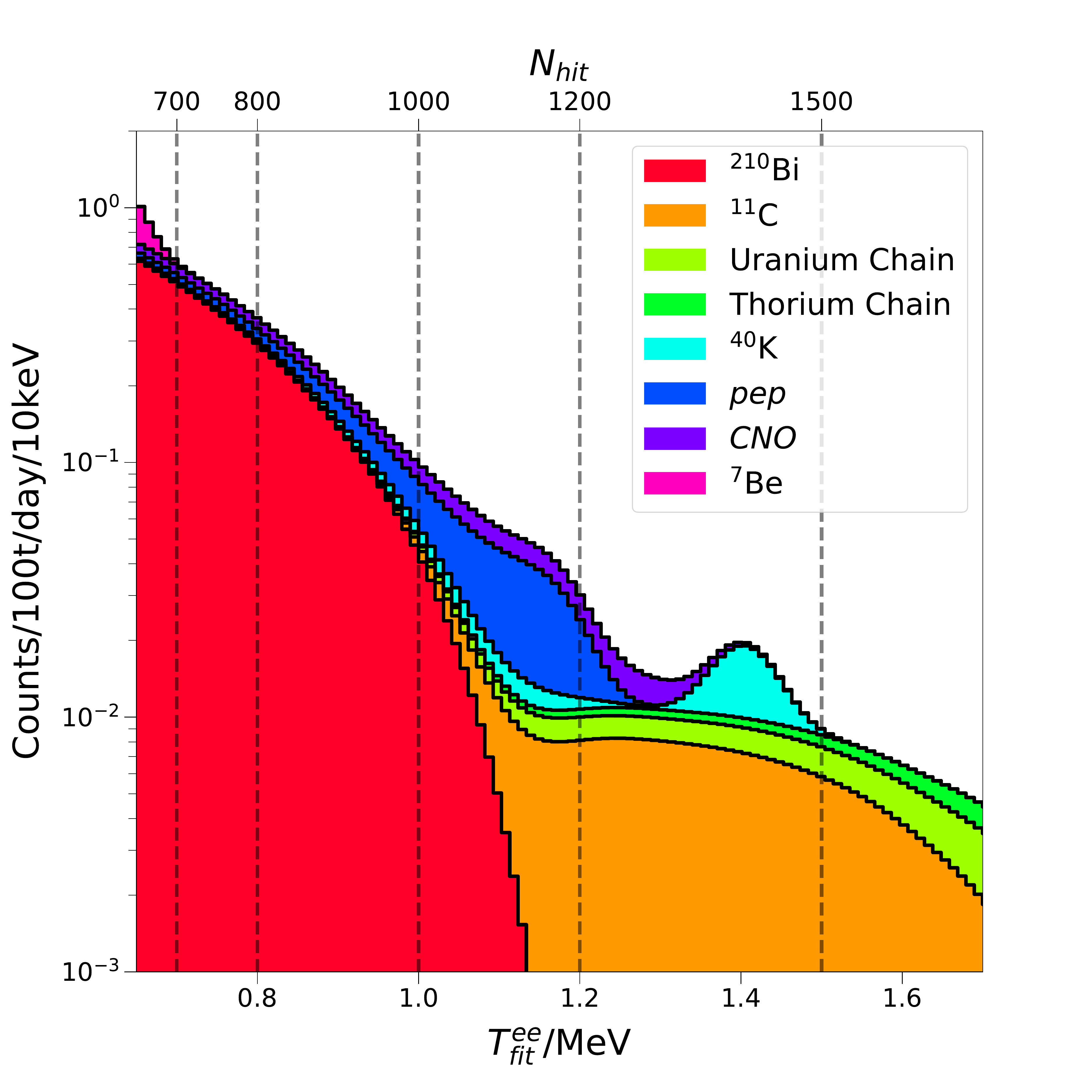} \\
    \includegraphics[width=6.5cm]{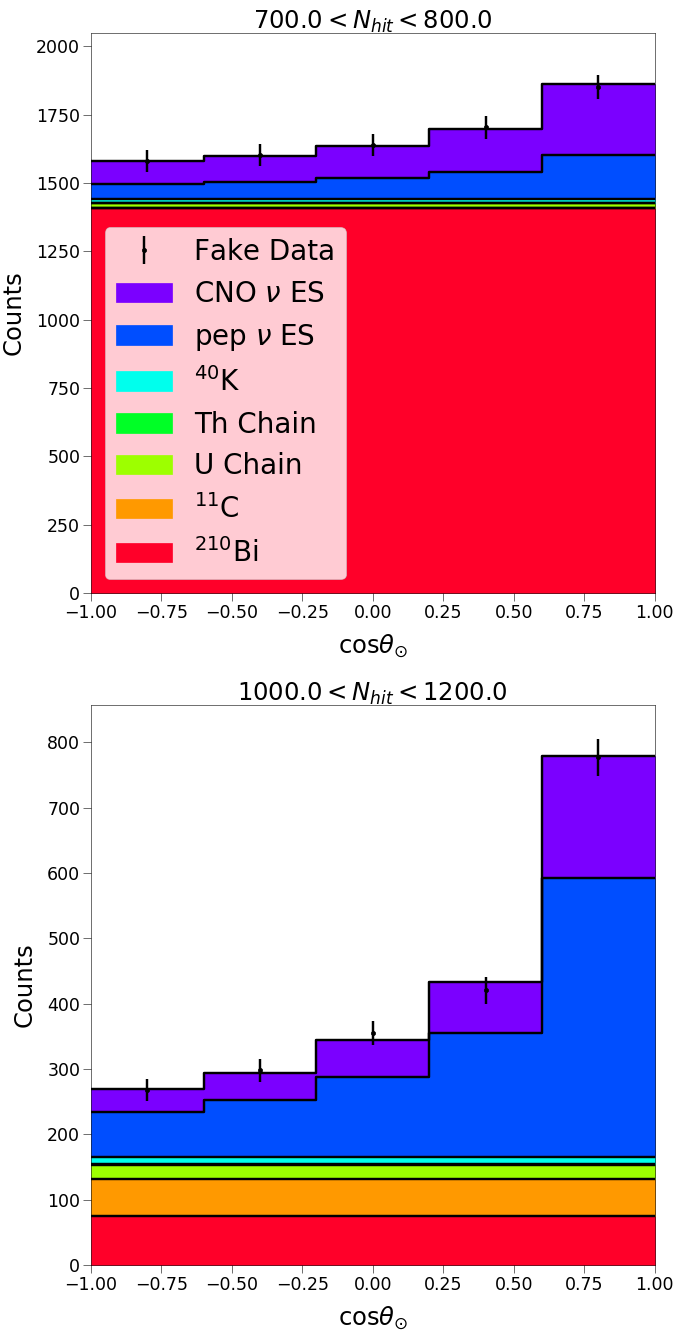}
    \end{tabular}
    \cprotect\caption{TOP: Background model for configurations \verb|77_FAST_1| and \verb|77_SLOW_1|. Dashed lines separate the four nhit bins used in CNO neutrino signal extraction. BOTTOM: an example fit for the \verb|77_FAST_1| configuration for two regions of $N_{hit}$ after an exposure of 500tyr.}
    \label{fig:CNO_background_model}
\end{figure}

To explore this capability, Azimov data sets \cite{Cowan:2010js} at a range of exposures were constructed for the background model shown in Figure~\ref{fig:CNO_background_model}, based on Borexino Phase I levels \cite{Bellini}, cosmogenic $^{11}$C levels at SNOLAB muon rates \cite{snolabscience}, and pep flux of $1.44 \times 10^8$ cm$^{-2}$ s$^{-1}$ \cite{Bergstrom}. The CNO event rates were calculated using predicted fluxes from the high metallicity Standard Solar Model \cite{cerdeno}.

\begin{figure}[H]
    \centering
    \includegraphics[width=\linewidth]{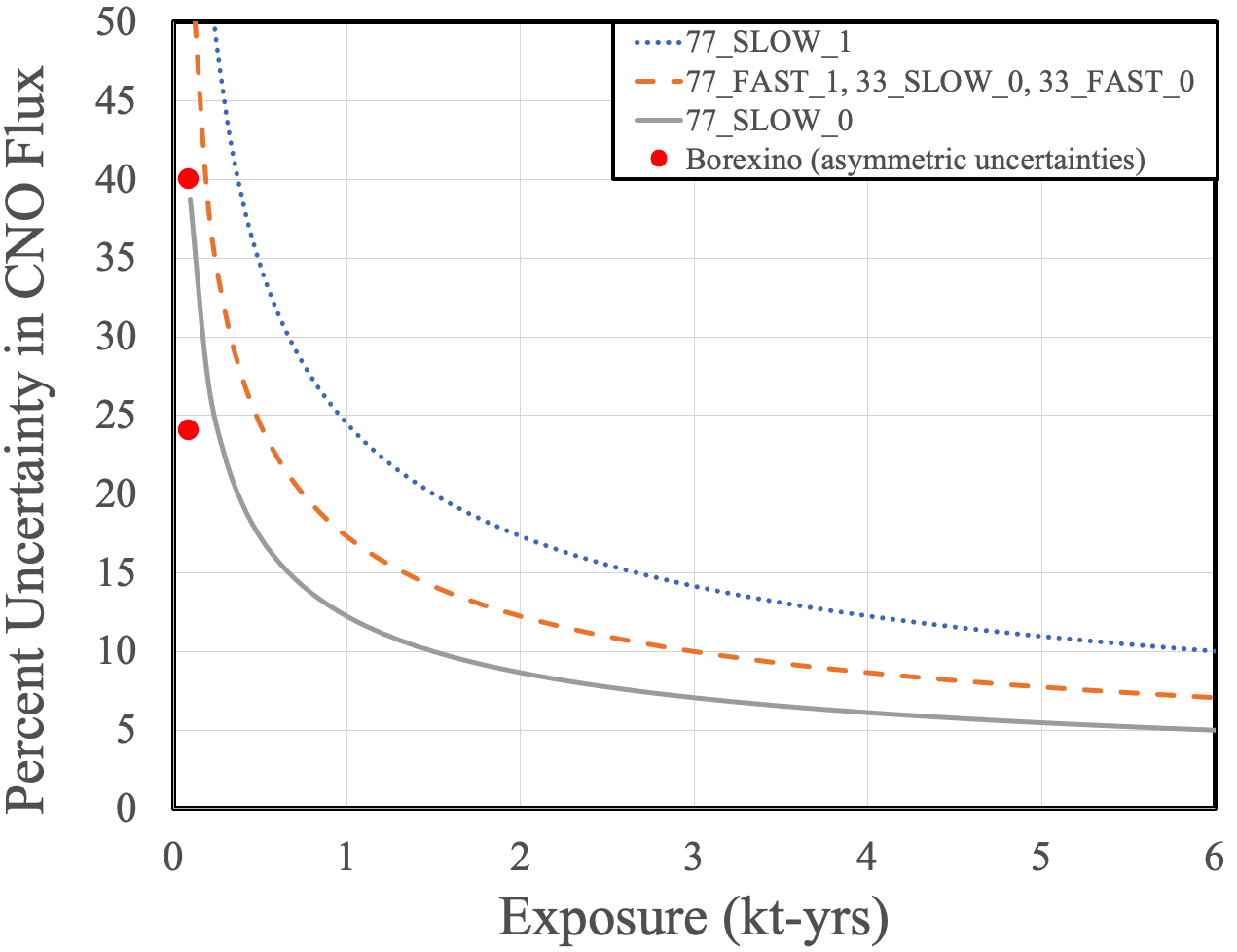}
    \cprotect\caption{Expected 1$\sigma$ fractional error on the CNO flux as a function of exposure for a range of detector configurations, assuming high metallicity. The asymmetric uncertainties from Borexino are also shown based on 4 live years of data \cite{Borexino_Nature}.}
    \label{fig:cno_sensitivity}
\end{figure}

The normalisations of the CNO signal and the backgrounds (including $^{210}$Bi), were then treated as free parameters in a Bayesian fit using a uniform, non-negative prior in the rate. The likelihood was modelled using 2D binned PDFs in $N_{hit}$ (figure~\ref{fig:CNO_background_model}) and reconstructed solar angle, $\cos\theta{\odot}$, estimated from a large number of simulated events for each detector configuration. Posterior samples from Markov chain Monte Carlo were used to estimate the most probable CNO normalisation and the corresponding shortest credible interval at each exposure. An example fit is shown in figure~\ref{fig:CNO_background_model} for two different $N_{hit}$ ranges. A combination of direct fits to different scenarios and scaling arguments were then used to project the expected fractional error on the CNO flux as a function of exposure for a range of detector configurations shown in Figure~\ref{fig:cno_sensitivity}. The results suggest that a detector with as little as $\sim$30\% coverage by standard HQE PMTs could be able to make a measurement of the CNO solar neutrino flux to a precision of better than 10\% with a few kiloton-years of exposure.

\section{\label{sec:neutrinoless_double_beta_decay} Neutrinoless Double Beta Decay}

An important area of application is the loading of liquid scintillators with isotopes, such as $^{130}$Te or $^{136}$Xe, to search for $0\nu\beta\beta$. As previously mentioned, the detected light output of acenapthene is $\sim$35\% lower than PPO, so the advantages of Cherenkov separation with this particular fluor choice needs to be balanced against the need for energy resolution. In addition, the light output of acenaphthene is particularly sensitive to quenching effects, so may be further impacted depending on the particular loading method. Other slow fluor combinations discussed in \cite{SlowFluors} may therefore be better choices for this application. Nevertheless, acenapthene will be used here to explore the in-principle characteristics of such an approach as a `best case' scenario for what might be achieved.

\subsection{\label{sec:bb_backgrounds} Backgrounds}

\subsubsection{\boroneight{} rejection}
\begin{figure}[H]
	\begin{subfigure}{1\columnwidth}
	  \centering
	  \includegraphics[width=0.9\linewidth]{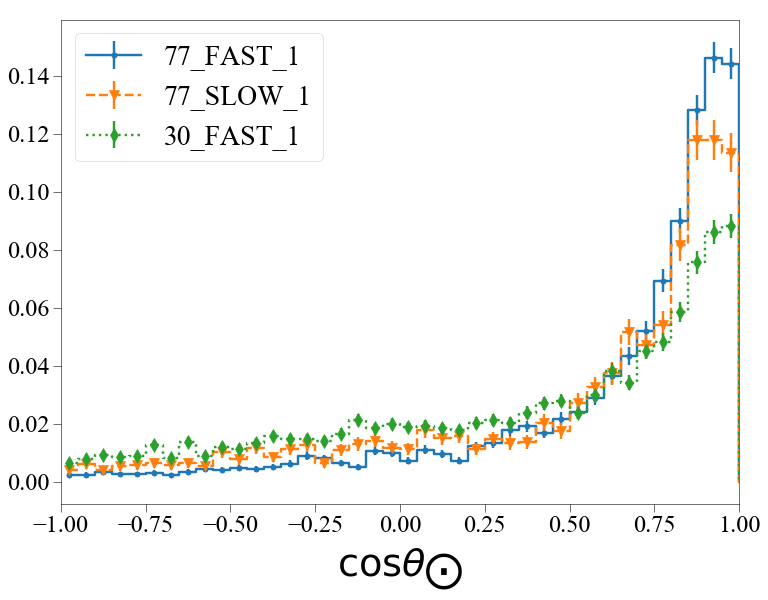}
	  \caption{}
	  \label{fig:solar_pointing_resolution}
    \end{subfigure}
      \newline
    \begin{subfigure}{1\columnwidth}
	  \centering
	  \includegraphics[width=0.9\linewidth]{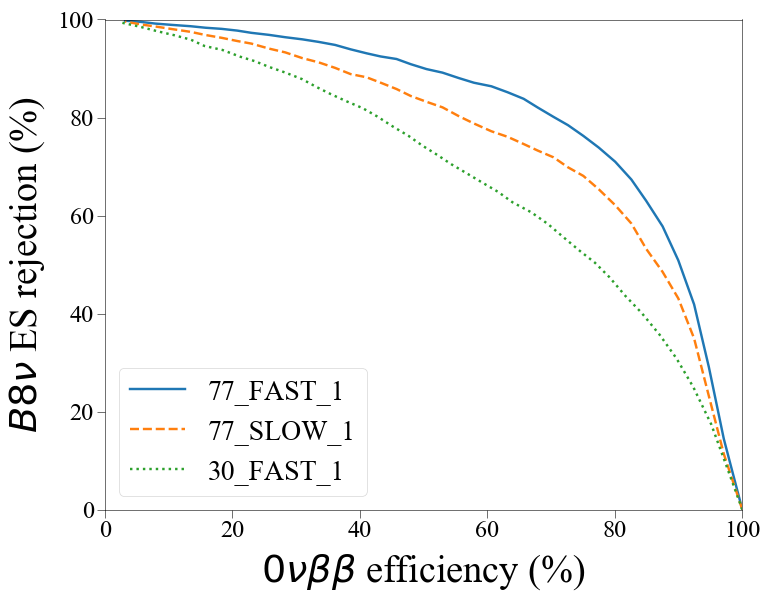}
	  \caption{}
      \label{fig:solar_cut_efficiency}
   \end{subfigure}
   \caption{(a): reconstructed solar direction for \boroneight{} neutrino events. (b): the rejection based on a simple directional cut on $\cos \theta_{\bigodot}$. Events cut to select reconstructed energies 2-3 MeV.}
    \label{fig:solar_rejection}
\end{figure}

Electrons produced through elastic scattering by \boroneight{} neutrino interactions will typically point away from the sun. With knowledge of the sun's position relative to the detector, it is therefore possible to reject \boroneight{} events that fall into the $0\nu\beta\beta$ energy window with a simple directional cut. As both $^{130}$Te and $^{136}$Xe have endpoints around 2.5 MeV, the reconstructed direction resolution and rejection vs efficiency curve for electrons in an energy window 2-3 MeV is given in Figure~\ref{fig:solar_rejection}. This suggests that a configuration with high PMT coverage might be able to reject $\sim$90\% of \boroneight{} events in the angular hemisphere defined by the solar direction, thus preserving half the $0\nu\beta\beta$ signal. This implies a gain in significance of $\sim 0.5/\sqrt{0.1} \sim 1.6$ if background levels are relatively high and \boroneight{} events are the dominant source. This therefore has the potential to be a very powerful technique, though not necessarily `transformational' in the presence of other backgrounds. 

\subsubsection{Cherenkov-based background rejection}
It has recently been pointed out \cite{Dunger:2019} that multi-site discrimination can be very effective in large scale liquid scintillators to distinguish a true $0\nu\beta\beta$ signal from radioactive backgrounds, including those due to cosmogenic activation. This capability is critical to be able to make a strong claim of discovery. As the multi-site technique relies on the vertex resolution from timing, the ability to distinguish such events may potentially be degraded with slow scintillators. On the other hand, the prompt Cherenkov signal signature for the various event classes is expected to be different and may compensate for the poorer vertex resolution. To explore this, a similar likelihood ratio approach was employed, using probability density functions based on the residual time distributions of PMT hits for the different event classes relative to the reconstructed vertex. Specifically,

\begin{equation}
    \Delta \log \mathcal{L} = \sum_{i=1}^{N_{hit}} \left[ \log P_S(t_{res}^i) - \log P_B(t_{res}^i) \right]
    \label{eq:likelihood_ratio}
\end{equation}

\noindent where $t^i_{res}$ is  the time  residual of  the ith  hit, $N_{hit}$ is the number of PMT hits in the event and $P_{S/B}$ is  the  probability  of  observing  a  hit  with  time  residual $t^i_{res}$ in  a  signal or background event, respectively. Here, the signal is taken to be the expectation for $0\nu\beta\beta$ events with an energy of 2.5 MeV and background is the expectation for the particularly troublesome radioactive decays of cosmogenic isotopes $^{60}$Co and $^{22}$Na.

\begin{figure}[H]
	  \centering
	  \includegraphics[width=0.9\linewidth]{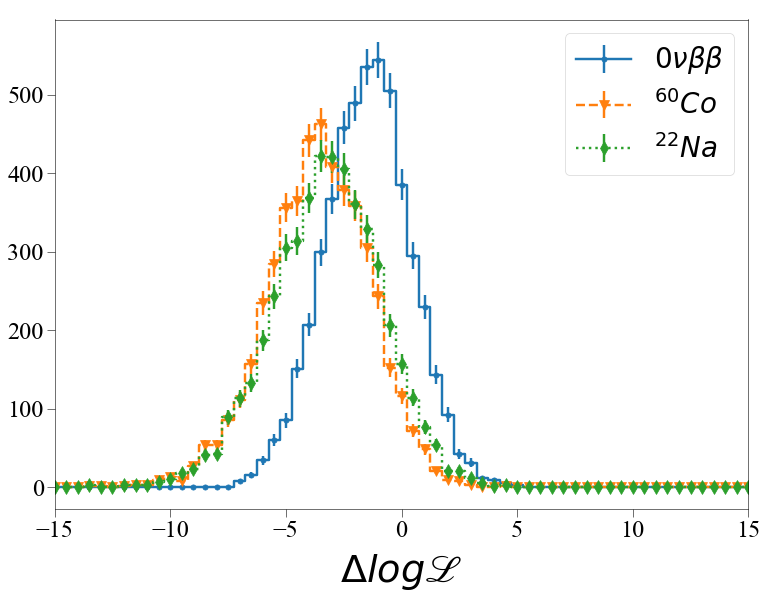}

   \cprotect\caption{$\Delta \log \mathcal{L}$ distributions for discriminating $0\nu\beta\beta$ from $^{60}$Co and $^{22}$Na for the \verb|77_FAST_1| configuration. The full time residual spectum was used in the likelihood calculation. Energy window = [2.4, 2.6] MeV.}
    \label{fig:0nu_Co60_rejection}
\end{figure}

The discrimination power against these backgrounds for detector configuration \verb|77_FAST_1| is given in Figure~\ref{fig:0nu_Co60_rejection}. For high photocathode coverage with relatively fast PMTs, the separation is slightly worse than shown in \cite{Dunger:2019} for typical fast scintillator and PMTs, but is still useful. Separations of this scale can be effective in breaking degeneracies to resolve the presence of such backgrounds in the context of a likelihood fit for the $0\nu\beta\beta$ signal.

\subsection{\label{sec:mechanism} Mechanism}
Once a $0\nu\beta\beta$ signal is established, the next question of high priority is to determine the underlying mechanism. Two of the most prominent models, left handed neutrino exchange (LNE) and right-handed current (RHC), can, in principle, be distinguished by the angular and energy distributions of the two emitted electrons. To date, only thin target tracking detectors such as NEMO and SuperNEMO \cite{nemo, supernemo} have sought to exploit this difference. Directional Cherenkov information, at least in principle, offers the possibility of an alternative route with a more scalable detector technology.

As an initial exploration of this capability in liquid scintillator using slow fluors, the presence of a highly significant $0\nu\beta\beta$ signal will be assumed. A likelihood-based
fit is then used to jointly estimate the directions of both electrons $\hat{d}_{1,2}$. The apparent separation angle between the two can be calculated using the dot product of these two directions:

\begin{equation}
    \cos \theta_{sep} = \hat{d}_1 \cdot \hat{d}_2
\end{equation}

\begin{figure}[H]
  \begin{subfigure}{0.9\columnwidth}
    \centering
    \includegraphics[width=1.\linewidth]{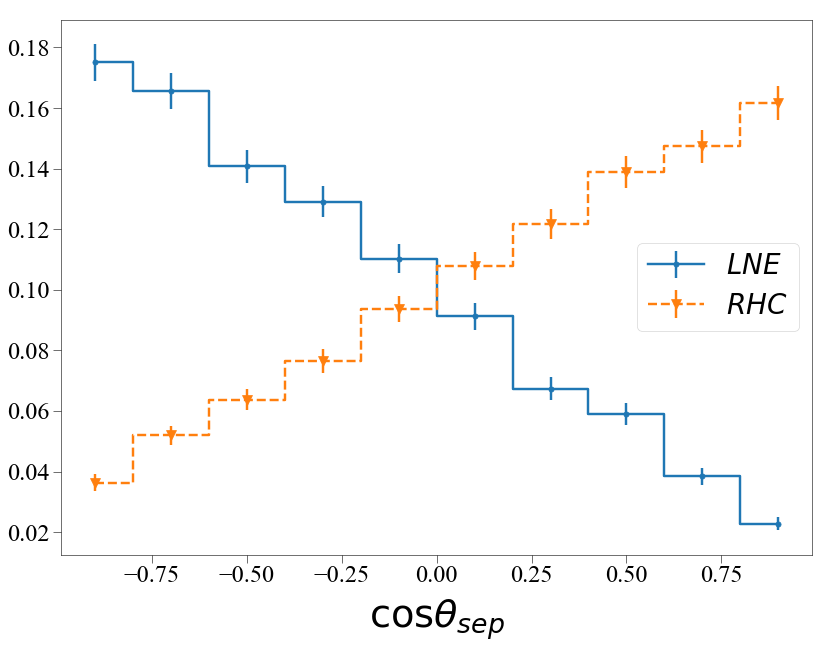}
    \caption{}
    \label{fig:0nubb_cosThetaSep_truth}
  \end{subfigure} \\
  \begin{subfigure}{0.9\columnwidth}
    \centering
    \includegraphics[width=1.\linewidth]{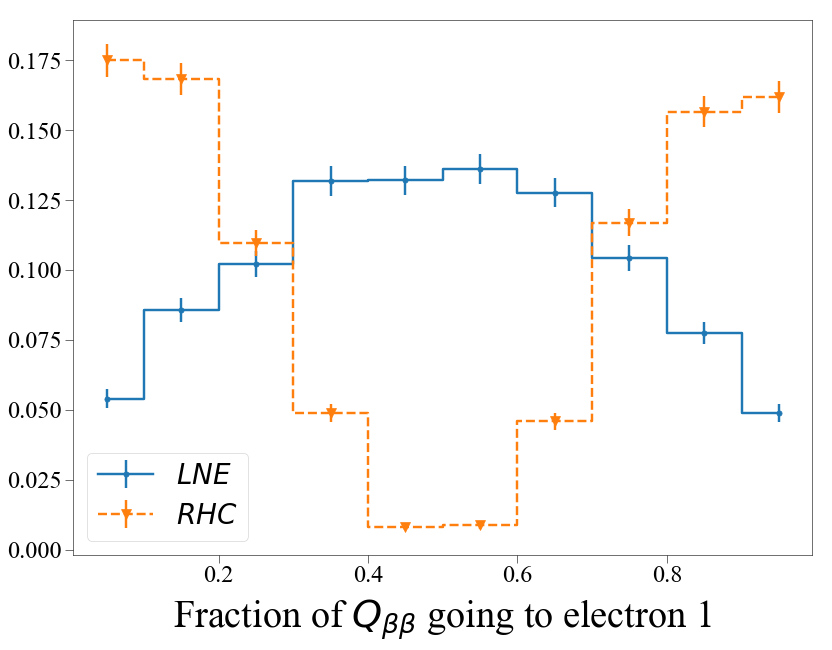}
    \caption{}
    \label{fig:0nubb_energy_frac_truth}
  \end{subfigure}
  \caption{Model dependence of $0\nu \beta \beta$ decay kinematics for LNE and RHC mechanisms.}
  \label{fig:0nubb_mechanism_truth}
\end{figure}

The fit also estimates the energy split between the two electrons, which is parameterised by the angular variable $\chi$ in order to avoid hard boundaries in the minimisation space. Thus, the electrons have kinetic energies given by: 

\begin{align}
    T_1 &= \sin^2 \chi \cdot Q_{\beta\beta} \\
    T_2 &= \cos^2 \chi \cdot Q_{\beta\beta}
\end{align}

\noindent where $Q_{\beta\beta}$ is 2.5~MeV. The true distributions ({\em i.e.} not based on reconstruction) of $\cos \theta_{sep}$ and $T_1$ are given in Figure~\ref{fig:0nubb_mechanism_truth}.

It is useful to combine this information into a single parameter that describes the associated momentum imbalance. The energy of each particle can be related to the momentum as:

\begin{equation}
    p = \sqrt{T(T + 2m_e)}
\end{equation}

\noindent in natural units, where $m_e$ is the electron mass. The momentum imbalance of the electron pair ($\mathcal{E}$) is defined as:

\begin{equation}
    \mathcal{E} = |p_1\hat{d}_1 + p_2\hat{d}_2 |
\end{equation}

\begin{figure}[htbp]
	\centering
  \begin{subfigure}{0.9\columnwidth}
    \centering
    \includegraphics[width=1.\linewidth]{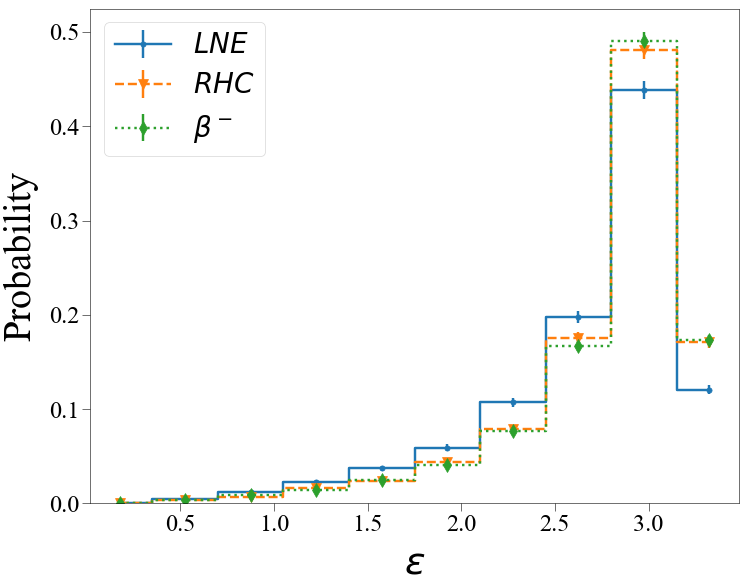}
    \caption{}
    \label{fig:0nubb_discriminants_1}
  \end{subfigure} \\
  \begin{subfigure}{0.85\columnwidth}
    \centering
    \includegraphics[width=1.\linewidth]{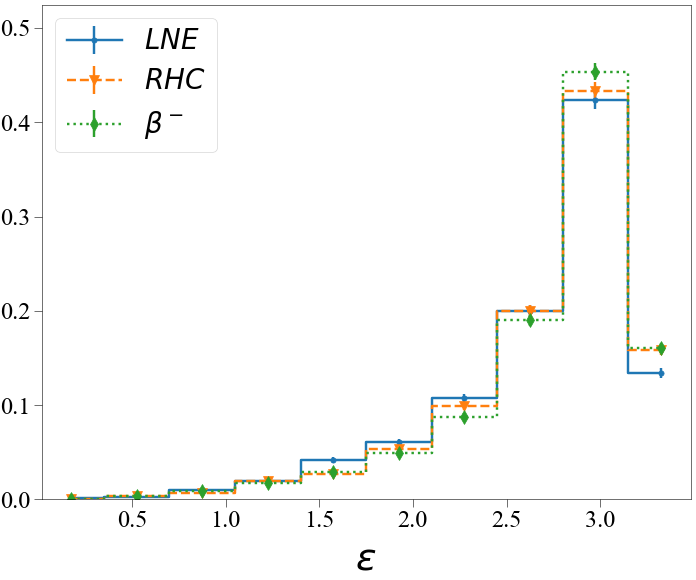}
    \caption{}
    \label{fig:0nubb_discriminants_2}
  \end{subfigure}

	\caption{Fitted values for $\mathcal{E}$ for LNE and RHC events, 2.5~MeV $e^-$ are shown for comparison. (a): minimizing with vertex time and position values fixed to the truth. (b): Floating all four parameters.}
	\label{fig:0nubb_discriminants}
\end{figure}

Figure~\ref{fig:0nubb_discriminants} shows 
the distribution of reconstructed values of $\mathcal{E}$ for $0\nu\beta\beta$ events for the Light Neutrino Exchange (LNE) and Right Handed Currents (RHC) mechanisms. Single 2.5~MeV electrons are also shown for reference.

Using the true vertex, the $\mathcal{E}$ distributions are as expected: single electrons produce the most imbalanced $\mathcal{E}$, followed by RHC events and then by LNE. However, using the reconstructed position, the differences between the distributions are notably degraded. 

To indicate the potential to discriminate between mechanisms, some number, $n_{obs}$, of pure $0\nu\beta\beta$ events are assumed to have been observed. We then define the log-likelihood difference between LNE and RHC hypotheses:

\begin{equation}
    \Delta \log \mathcal{L} = \sum_{i=1}^{n_{obs}} \left[ \log P_{LNE}\{\mathcal{E}^i\} - \log P_{RHC}\{\mathcal{E}^i\}
    \right]
\end{equation}

\begin{figure}[H]
	\centering
	\includegraphics[width=0.48\textwidth]{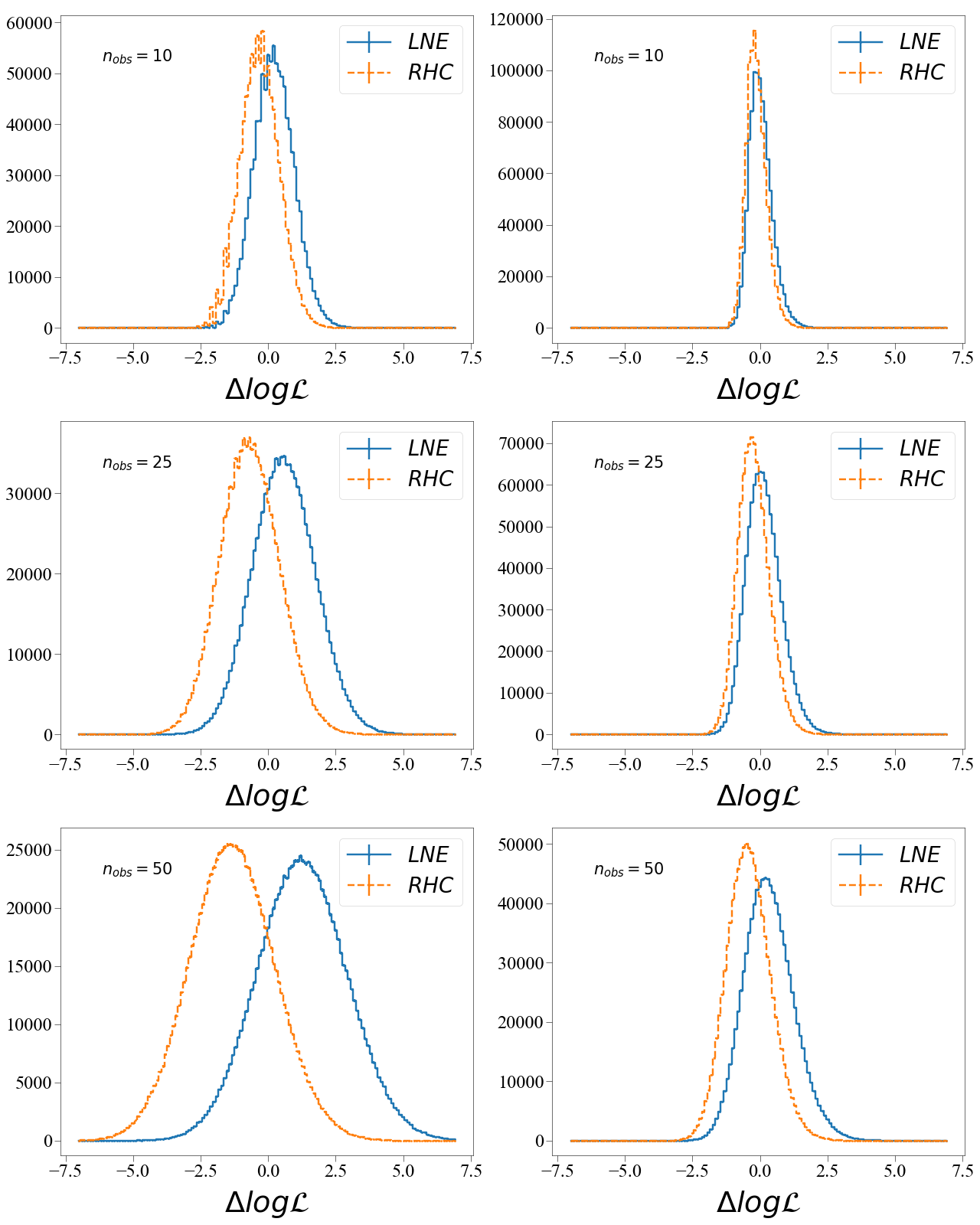}
	\caption{$\Delta \log \mathcal{L}$ distributions for LNE and RHC mechanisms for different assumed numbers of $0\nu\beta\beta$ events. Left: fixing the true vertex time and position. Right: floating the vertex time and position in the fit.}
	\label{fig:0nubb_dll_distributions}
\end{figure}

\begin{figure}[H]
	\centering
	\includegraphics[width=0.48\textwidth]{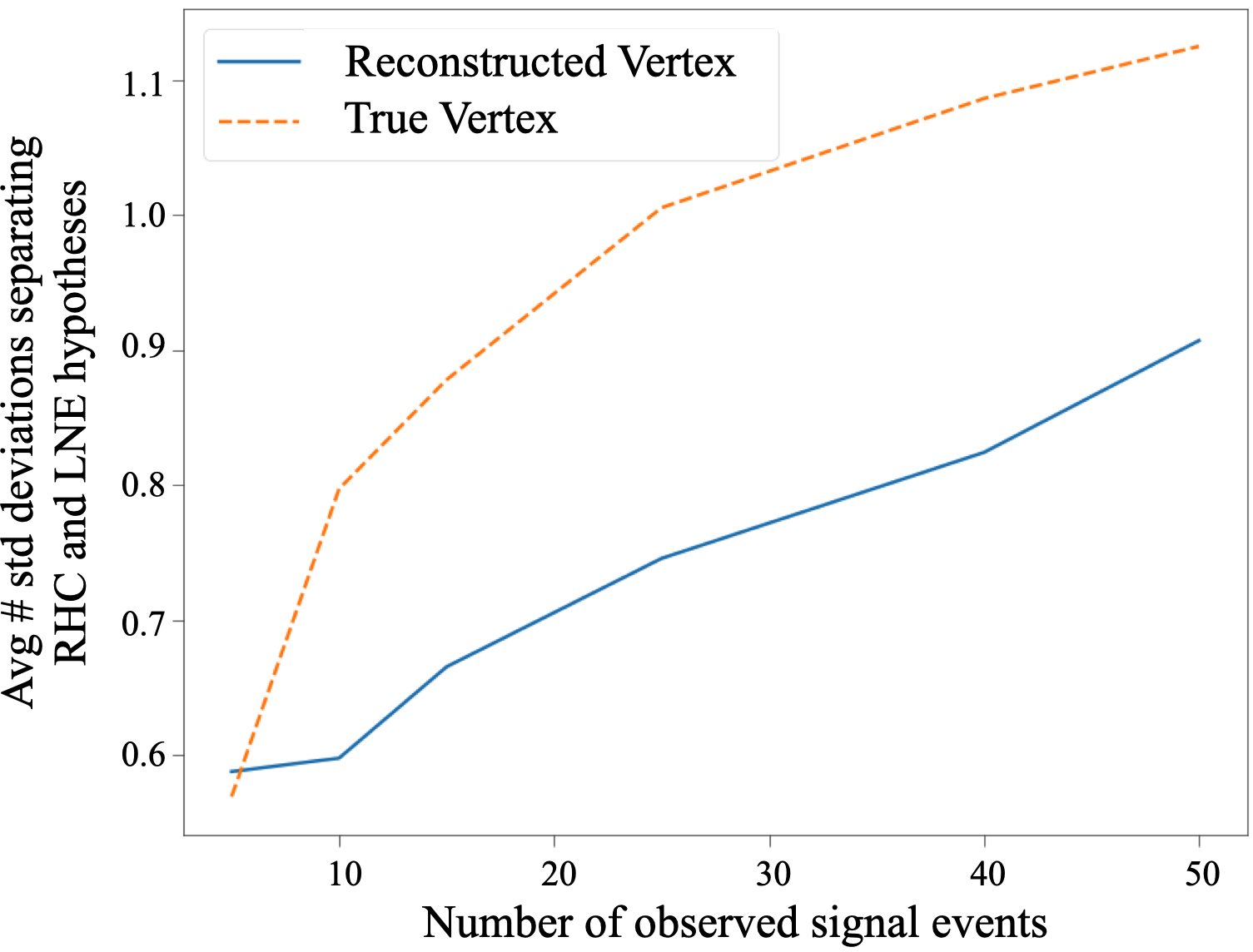}
	\caption{Predicted average number of standard deviations separating RHC from LNE mechanisms as a function of the number of signal events observed assuming the true vertex (upper curve) and reconstructed vertex (lower curve) positions.}
	\label{fig:mechanism_sensitivity}
\end{figure}

The distribution of $\Delta \log \mathcal{L}$ has been estimated with a toy Monte Carlo for a variety of different $n_{obs}$ for both mechanisms (Figure~\ref{fig:0nubb_dll_distributions}). This is translated into the number of standard deviations separating the mechanisms as a function of observed signal events in Figure~\ref{fig:mechanism_sensitivity}. Only weak discrimination is observed, particularly when using reconstructed vertex positions.

\section{\label{sec:conclusions} Conclusions}
Simulations of a large-scale liquid scintillator detector of conventional construction and based on LAB, but using acenaphthene as a primary slow fluor, have been used to explore the physics potential for measurements of low energy CNO solar neutrinos and $0\nu\beta\beta$. For CNO neutrinos, directional information from the time-separated Cherenkov light is predicted to be good enough to break important degeneracies between signal and background to a high degree, including those from the particularly problematic $^{210}$Bi. The most efficient configuration appears to be the use of acenaphthene by itself, without a secondary fluor, which both increases the number of unscattered Cherenkov photons detected and further reducing the scintillation component, while still maintaining sufficient vertex and energy resolution. These studies suggest that a detector with as little as $\sim$30\% photocathode coverage with commercially available HQE PMTs should be able to determine the CNO flux to a precision of better than 10\% with just a few kiloton-years of exposure, assuming ambient background levels comparable with that achieved by Borexino. 

For $0\nu\beta\beta$, solar neutrinos themselves form an important component of backgrounds, and direction information from the time-separated Cherenkov light can thus be used to identify and reduce this background. The choice of acenaphthene may not necessarily be optimal, depending on the isotope loading technique and levels of other backgrounds, but is used here as a best-case scenario given its effectiveness for separating Cherenkov light. We find that backgrounds from solar neutrinos can potentially be suppressed by a factor of $\sim$10 in the angular hemisphere defined by the solar direction, improving the $signal/\sqrt{background}$ by $\sim$1.6 where backgrounds are relatively high and dominated by solar neutrinos. In principle, given a strong enough $0\nu\beta\beta$ signal, one could also use angular information from Cherenkov light to distinguish between LNE and RHC mechanisms. However, this appears to be a daunting prospect in practice, particularly for slow fluors, where vertex resolution is compromised. Other approaches that seek to separate Cherenkov light while maintaining fast timing (such as the use of dichroic surfaces) might be worth exploring, but would have to achieve very high coverage for that Cherenkov light detection without sacrificing energy resolution through reduced collection of scintillation photons.

\section{Acknowledgements}
We wish to acknowledge members of the SNO+ collaboration for useful discussions and, in particular, to thank Rehan Deen at Oxford for earlier exploration of extracting event characteristics with slow fluors. The work of the authors has been supported by the Science Technology and Facilities Council (STFC) of the United Kingdom, grant number ST/S000933/1. Appropriate representations of the data relevant to the conclusions have been provided within this paper. For the purposes of open access, the authors have applied a Creative Commons Attribution licence to any Author Accepted Manuscript version arising. 

\bibliography{slowscint}

\begin{thebibliography}{18}%
\makeatletter
\providecommand \@ifxundefined [1]{%
 \@ifx{#1\undefined}
}%
\providecommand \@ifnum [1]{%
 \ifnum #1\expandafter \@firstoftwo
 \else \expandafter \@secondoftwo
 \fi
}%
\providecommand \@ifx [1]{%
 \ifx #1\expandafter \@firstoftwo
 \else \expandafter \@secondoftwo
 \fi
}%
\providecommand \natexlab [1]{#1}%
\providecommand \enquote  [1]{``#1''}%
\providecommand \bibnamefont  [1]{#1}%
\providecommand \bibfnamefont [1]{#1}%
\providecommand \citenamefont [1]{#1}%
\providecommand \href@noop [0]{\@secondoftwo}%
\providecommand \href [0]{\begingroup \@sanitize@url \@href}%
\providecommand \@href[1]{\@@startlink{#1}\@@href}%
\providecommand \@@href[1]{\endgroup#1\@@endlink}%
\providecommand \@sanitize@url [0]{\catcode `\\12\catcode `\$12\catcode
  `\&12\catcode `\#12\catcode `\^12\catcode `\_12\catcode `\%12\relax}%
\providecommand \@@startlink[1]{}%
\providecommand \@@endlink[0]{}%
\providecommand \url  [0]{\begingroup\@sanitize@url \@url }%
\providecommand \@url [1]{\endgroup\@href {#1}{\urlprefix }}%
\providecommand \urlprefix  [0]{URL }%
\providecommand \Eprint [0]{\href }%
\providecommand \doibase [0]{https://doi.org/}%
\providecommand \selectlanguage [0]{\@gobble}%
\providecommand \bibinfo  [0]{\@secondoftwo}%
\providecommand \bibfield  [0]{\@secondoftwo}%
\providecommand \translation [1]{[#1]}%
\providecommand \BibitemOpen [0]{}%
\providecommand \bibitemStop [0]{}%
\providecommand \bibitemNoStop [0]{.\EOS\space}%
\providecommand \EOS [0]{\spacefactor3000\relax}%
\providecommand \BibitemShut  [1]{\csname bibitem#1\endcsname}%
\let\auto@bib@innerbib\@empty
\bibitem [{\citenamefont {{Biller}}\ \emph {et~al.}(2020)\citenamefont
  {{Biller}}, \citenamefont {{Leming}},\ and\ \citenamefont
  {{Paton}}}]{SlowFluors}%
  \BibitemOpen
  \bibfield  {author} {\bibinfo {author} {\bibfnamefont {S.~D.}\ \bibnamefont
  {{Biller}}}, \bibinfo {author} {\bibfnamefont {E.~J.}\ \bibnamefont
  {{Leming}}},\ and\ \bibinfo {author} {\bibfnamefont {J.~L.}\ \bibnamefont
  {{Paton}}},\ }\href {https://doi.org/10.1016/j.nima.2020.164106} {\bibfield
  {journal} {\bibinfo  {journal} {Nuclear Instruments and Methods in Physics
  Research A}\ }\textbf {\bibinfo {volume} {972}},\ \bibinfo {eid} {164106}
  (\bibinfo {year} {2020})},\ \Eprint {https://arxiv.org/abs/2001.10825}
  {arXiv:2001.10825 [physics.ins-det]} \BibitemShut {NoStop}%
\bibitem [{\citenamefont {Land}\ \emph {et~al.}(2021)\citenamefont {Land},
  \citenamefont {Bagdasarian}, \citenamefont {Caravaca}, \citenamefont
  {Smiley}, \citenamefont {Yeh},\ and\ \citenamefont {Orebi~Gann}}]{Land}%
  \BibitemOpen
  \bibfield  {author} {\bibinfo {author} {\bibfnamefont {B.~J.}\ \bibnamefont
  {Land}}, \bibinfo {author} {\bibfnamefont {Z.}~\bibnamefont {Bagdasarian}},
  \bibinfo {author} {\bibfnamefont {J.}~\bibnamefont {Caravaca}}, \bibinfo
  {author} {\bibfnamefont {M.}~\bibnamefont {Smiley}}, \bibinfo {author}
  {\bibfnamefont {M.}~\bibnamefont {Yeh}},\ and\ \bibinfo {author}
  {\bibfnamefont {G.~D.}\ \bibnamefont {Orebi~Gann}},\ }\href
  {https://doi.org/10.1103/PhysRevD.103.052004} {\bibfield  {journal} {\bibinfo
   {journal} {Phys. Rev. D}\ }\textbf {\bibinfo {volume} {103}},\ \bibinfo
  {pages} {052004} (\bibinfo {year} {2021})}\BibitemShut {NoStop}%
\bibitem [{\citenamefont {{T. Bolton et al.}}(2019)}]{rat}%
  \BibitemOpen
  \bibfield  {author} {\bibinfo {author} {\bibnamefont {{T. Bolton et al.}}},\
  }\href@noop {} {\bibinfo {title} {{RAT} users{{'}} guide,
  \url{https://rat.readthedocs.io/en/latest/}}} (\bibinfo {year} {accessed on
  Janurary 1st 2019})\BibitemShut {NoStop}%
\bibitem [{\citenamefont {{G. A Horton-Smith}}(2019)}]{glg4sim}%
  \BibitemOpen
  \bibfield  {author} {\bibinfo {author} {\bibnamefont {{G. A Horton-Smith}}},\
  }\href@noop {} {\bibinfo {title} {{Generic liquid scintillator {G}eant4
  simulation}, \url{http://neutrino.phys.ksu.edu/∼GLG4sim/}}} (\bibinfo
  {year} {accessed on Janurary 1st 2019})\BibitemShut {NoStop}%
\bibitem [{\citenamefont {{S. Agostinelli et al.}}(2003)}]{geant4}%
  \BibitemOpen
  \bibfield  {author} {\bibinfo {author} {\bibnamefont {{S. Agostinelli et
  al.}}},\ }\href
  {https://doi.org/https://doi.org/10.1016/S0168-9002(03)01368-8} {\bibfield
  {journal} {\bibinfo  {journal} {Nucl. Instrum. Meth.}\ }\textbf {\bibinfo
  {volume} {A506}},\ \bibinfo {pages} {250 } (\bibinfo {year}
  {2003})}\BibitemShut {NoStop}%
\bibitem [{\citenamefont {Boger}\ \emph {et~al.}(2000)\citenamefont {Boger}
  \emph {et~al.}}]{Boger:1999bb}%
  \BibitemOpen
  \bibfield  {author} {\bibinfo {author} {\bibfnamefont {J.}~\bibnamefont
  {Boger}} \emph {et~al.} (\bibinfo {collaboration} {SNO}),\ }\href
  {https://doi.org/10.1016/S0168-9002(99)01469-2} {\bibfield  {journal}
  {\bibinfo  {journal} {Nucl. Instrum. Meth.}\ }\textbf {\bibinfo {volume}
  {A449}},\ \bibinfo {pages} {172} (\bibinfo {year} {2000})},\ \Eprint
  {https://arxiv.org/abs/nucl-ex/9910016} {arXiv:nucl-ex/9910016 [nucl-ex]}
  \BibitemShut {NoStop}%
\bibitem [{\citenamefont {Ponkratenko}\ \emph {et~al.}(2000)\citenamefont
  {Ponkratenko}, \citenamefont {Tretyak},\ and\ \citenamefont
  {Zdesenko}}]{Ponkratenko:2000um}%
  \BibitemOpen
  \bibfield  {author} {\bibinfo {author} {\bibfnamefont {O.~A.}\ \bibnamefont
  {Ponkratenko}}, \bibinfo {author} {\bibfnamefont {V.~I.}\ \bibnamefont
  {Tretyak}},\ and\ \bibinfo {author} {\bibfnamefont {{\relax Yu}.~G.}\
  \bibnamefont {Zdesenko}},\ }\href {https://doi.org/10.1134/1.855784}
  {\bibfield  {journal} {\bibinfo  {journal} {Phys. Atom. Nucl.}\ }\textbf
  {\bibinfo {volume} {63}},\ \bibinfo {pages} {1282} (\bibinfo {year}
  {2000})},\ \bibinfo {note} {[Yad. Fiz.63,1355(2000)]},\ \Eprint
  {https://arxiv.org/abs/nucl-ex/0104018} {arXiv:nucl-ex/0104018 [nucl-ex]}
  \BibitemShut {NoStop}%
\bibitem [{\citenamefont {An}\ \emph {et~al.}(2016)\citenamefont {An} \emph
  {et~al.}}]{JUNO}%
  \BibitemOpen
  \bibfield  {author} {\bibinfo {author} {\bibfnamefont {F.}~\bibnamefont {An}}
  \emph {et~al.} (\bibinfo {collaboration} {JUNO}),\ }\href
  {https://doi.org/10.1088/0954-3899/43/3/030401} {\bibfield  {journal}
  {\bibinfo  {journal} {J. Phys. G}\ }\textbf {\bibinfo {volume} {43}},\
  \bibinfo {pages} {030401} (\bibinfo {year} {2016})},\ \Eprint
  {https://arxiv.org/abs/1507.05613} {arXiv:1507.05613 [physics.ins-det]}
  \BibitemShut {NoStop}%
\bibitem [{\citenamefont {{M. Agostini et al.}}(2020)}]{Borexino_Nature}%
  \BibitemOpen
  \bibfield  {author} {\bibinfo {author} {\bibnamefont {{M. Agostini et
  al.}}},\ }\href {https://doi.org/https://doi.org/10.1038/s41586-020-2934-0}
  {\bibfield  {journal} {\bibinfo  {journal} {Nature}\ }\textbf {\bibinfo
  {volume} {587}},\ \bibinfo {pages} {577 } (\bibinfo {year}
  {2020})}\BibitemShut {NoStop}%
\bibitem [{\citenamefont {{M. Agostini et al.}}(2021)}]{Borexino_direction}%
  \BibitemOpen
  \bibfield  {author} {\bibinfo {author} {\bibnamefont {{M. Agostini et
  al.}}},\ }\href@noop {} {\bibfield  {journal} {\bibinfo  {journal}
  {arXiv:2112.11816 [hep-ex]}\ } (\bibinfo {year} {2021})}\BibitemShut
  {NoStop}%
\bibitem [{\citenamefont {et~al.}(2011)}]{Cowan:2010js}%
  \BibitemOpen
  \bibfield  {author} {\bibinfo {author} {\bibfnamefont {G.~C.}\ \bibnamefont
  {et~al.}},\ }\href {https://doi.org/10.1140/epjc/s10052-011-1554-0,
  10.1140/epjc/s10052-013-2501-z} {\bibfield  {journal} {\bibinfo  {journal}
  {Eur. Phys. J.}\ }\textbf {\bibinfo {volume} {C71}},\ \bibinfo {pages} {1554}
  (\bibinfo {year} {2011})},\ \bibinfo {note} {[Erratum: Eur. Phys.
  J.C73,2501(2013)]},\ \Eprint {https://arxiv.org/abs/1007.1727}
  {arXiv:1007.1727 [physics.data-an]} \BibitemShut {NoStop}%
\bibitem [{\citenamefont {Bellini}\ \emph {et~al.}(2014)\citenamefont {Bellini}
  \emph {et~al.}}]{Bellini}%
  \BibitemOpen
  \bibfield  {author} {\bibinfo {author} {\bibfnamefont {G.}~\bibnamefont
  {Bellini}} \emph {et~al.} (\bibinfo {collaboration} {Borexino}),\ }\href
  {https://doi.org/10.1103/PhysRevD.89.112007} {\bibfield  {journal} {\bibinfo
  {journal} {Phys. Rev. D}\ }\textbf {\bibinfo {volume} {89}},\ \bibinfo
  {pages} {112007} (\bibinfo {year} {2014})},\ \Eprint
  {https://arxiv.org/abs/1308.0443} {arXiv:1308.0443 [hep-ex]} \BibitemShut
  {NoStop}%
\bibitem [{\citenamefont {Jillings}(2009)}]{snolabscience}%
  \BibitemOpen
  \bibfield  {author} {\bibinfo {author} {\bibfnamefont {C.}~\bibnamefont
  {Jillings}},\ }\href
  {https://doi.org/https://doi.org/10.1016/j.nuclphysbps.2009.02.031}
  {\bibfield  {journal} {\bibinfo  {journal} {Nuclear Physics B - Proceedings
  Supplements}\ }\textbf {\bibinfo {volume} {188}},\ \bibinfo {pages} {130}
  (\bibinfo {year} {2009})},\ \bibinfo {note} {proceedings of the Neutrino
  Oscillation Workshop}\BibitemShut {NoStop}%
\bibitem [{\citenamefont {Bergstrom}\ \emph {et~al.}(2016)\citenamefont
  {Bergstrom}, \citenamefont {Gonzalez-Garcia}, \citenamefont {Maltoni},
  \citenamefont {Pena-Garay}, \citenamefont {Serenelli},\ and\ \citenamefont
  {Song}}]{Bergstrom}%
  \BibitemOpen
  \bibfield  {author} {\bibinfo {author} {\bibfnamefont {J.}~\bibnamefont
  {Bergstrom}}, \bibinfo {author} {\bibfnamefont {M.~C.}\ \bibnamefont
  {Gonzalez-Garcia}}, \bibinfo {author} {\bibfnamefont {M.}~\bibnamefont
  {Maltoni}}, \bibinfo {author} {\bibfnamefont {C.}~\bibnamefont {Pena-Garay}},
  \bibinfo {author} {\bibfnamefont {A.~M.}\ \bibnamefont {Serenelli}},\ and\
  \bibinfo {author} {\bibfnamefont {N.}~\bibnamefont {Song}},\ }\href
  {https://doi.org/10.1007/JHEP03(2016)132} {\bibfield  {journal} {\bibinfo
  {journal} {JHEP}\ }\textbf {\bibinfo {volume} {03}}\bibfield  {number}
  {\bibinfo  {number} { (132)}},\ }\Eprint {https://arxiv.org/abs/1601.00972}
  {arXiv:1601.00972 [hep-ph]} \BibitemShut {NoStop}%
\bibitem [{\citenamefont {Cerdeno}\ \emph {et~al.}(2018)\citenamefont
  {Cerdeno}, \citenamefont {Davis}, \citenamefont {Fairbairn},\ and\
  \citenamefont {Vincent}}]{cerdeno}%
  \BibitemOpen
  \bibfield  {author} {\bibinfo {author} {\bibfnamefont {D.~G.}\ \bibnamefont
  {Cerdeno}}, \bibinfo {author} {\bibfnamefont {J.~H.}\ \bibnamefont {Davis}},
  \bibinfo {author} {\bibfnamefont {M.}~\bibnamefont {Fairbairn}},\ and\
  \bibinfo {author} {\bibfnamefont {A.~C.}\ \bibnamefont {Vincent}},\ }\href
  {https://doi.org/10.1088/1475-7516/2018/04/037} {\bibfield  {journal}
  {\bibinfo  {journal} {JCAP}\ }\textbf {\bibinfo {volume} {04}}\bibfield
  {number} {\bibinfo  {number} { (037)}},\ }\Eprint
  {https://arxiv.org/abs/1712.06522} {arXiv:1712.06522 [hep-ph]} \BibitemShut
  {NoStop}%
\bibitem [{\citenamefont {Dunger}\ and\ \citenamefont
  {Biller}(2019)}]{Dunger:2019}%
  \BibitemOpen
  \bibfield  {author} {\bibinfo {author} {\bibfnamefont {J.}~\bibnamefont
  {Dunger}}\ and\ \bibinfo {author} {\bibfnamefont {S.~D.}\ \bibnamefont
  {Biller}},\ }\href
  {https://doi.org/https://doi.org/10.1016/j.nima.2019.162420} {\bibfield
  {journal} {\bibinfo  {journal} {Nuclear Instruments and Methods in Physics
  Research Section A: Accelerators, Spectrometers, Detectors and Associated
  Equipment}\ }\textbf {\bibinfo {volume} {943}},\ \bibinfo {pages} {162420}
  (\bibinfo {year} {2019})}\BibitemShut {NoStop}%
\bibitem [{\citenamefont {Arnold}\ \emph {et~al.}(2011)\citenamefont {Arnold}
  \emph {et~al.}}]{nemo}%
  \BibitemOpen
  \bibfield  {author} {\bibinfo {author} {\bibfnamefont {R.}~\bibnamefont
  {Arnold}} \emph {et~al.} (\bibinfo {collaboration} {NEMO-3}),\ }\href
  {https://doi.org/10.1103/PhysRevLett.107.062504} {\bibfield  {journal}
  {\bibinfo  {journal} {Phys. Rev. Lett.}\ }\textbf {\bibinfo {volume} {107}},\
  \bibinfo {pages} {062504} (\bibinfo {year} {2011})},\ \Eprint
  {https://arxiv.org/abs/1104.3716} {arXiv:1104.3716 [nucl-ex]} \BibitemShut
  {NoStop}%
\bibitem [{\citenamefont {Arnold}\ \emph {et~al.}(2010)\citenamefont {Arnold}
  \emph {et~al.}}]{supernemo}%
  \BibitemOpen
  \bibfield  {author} {\bibinfo {author} {\bibfnamefont {R.}~\bibnamefont
  {Arnold}} \emph {et~al.} (\bibinfo {collaboration} {SuperNEMO}),\ }\href
  {https://doi.org/10.1140/epjc/s10052-010-1481-5} {\bibfield  {journal}
  {\bibinfo  {journal} {Eur. Phys. J. C}\ }\textbf {\bibinfo {volume} {70}},\
  \bibinfo {pages} {927} (\bibinfo {year} {2010})},\ \Eprint
  {https://arxiv.org/abs/1005.1241} {arXiv:1005.1241 [hep-ex]} \BibitemShut
  {NoStop}%
\end{thebibliography}%

\end{document}